\documentclass[aps,
    pra,
    twocolumn, 
    longbibliography=false]{revtex4-2}

\usepackage{natbib}
\usepackage{tikz}
\usepackage{pgfplots}
\pgfplotsset{compat=1.18}
\usepackage{sansmath} 
\pgfplotsset{
    every axis/.append style={
        tick label style={font=\sansmath\sffamily},
        label style={font=\sansmath\sffamily},
        legend style={font=\sansmath\sffamily},
    }
}
\usepgfplotslibrary{groupplots}
\usepackage{siunitx}
\usepackage[english]{babel}
\usepackage{amsmath,amssymb,mathrsfs}
\usepackage{psfrag}
\usepackage{graphicx}
\usepackage{graphics}
\usepackage{epsfig}
\usepackage{bm}
\usepackage{color}
\usepackage{verbatim,color}
\usepackage{physics}
\usepackage[normalem]{ulem}
\usepackage{cancel}
\usepackage[colorlinks=true,linkcolor=blue,citecolor=black,urlcolor=blue]{hyperref}

\usepackage[acronym]{glossaries}
\newacronym{dr}{DR}{dimensional regularization}
\newacronym{ms}{MS}{minimal subtraction}
\newacronym{eft}{EFT}{effective field theory}
\newacronym{mqst}{MQST}{macroscopic quantum self-trapping }
\newacronym{lhy}{LHY}{Lee-Huang-Yang}
\newcommand{\beq}{\begin{equation}}
\newcommand{\eeq}{\end{equation}}
\newcommand{\beqa}{\begin{eqnarray}}
\newcommand{\eeqa}{\end{eqnarray}}
\newcommand{\ba}{\begin{aligned}[b]}
\newcommand{\ea}{\end{aligned}}

\definecolor{darkgreen}{rgb}{0.0, 0.5, 0.0}
\definecolor{shadow}{rgb}{0.6, 0.6, 0.6}
\definecolor{darkgreen}{rgb}{0.0, 0.5, 0.0}
\definecolor{c1}{rgb}{0.9, 0.0, 0.0}
\definecolor{c2}{rgb}{0.3, 0.0, 0.6}
\definecolor{c3}{rgb}{0.9, 0.0, 0.0}

\newcommand{\rev}[1]{#1}

\newcommand{\mn}[1]{\marginpar{\tiny}}

\usepackage{ragged2e} 
\makeatletter
\long\def\@makecaption#1#2{%
  \vskip\abovecaptionskip
  \parbox{\linewidth}{\justifying #1\ #2\par}%
  \vskip\belowcaptionskip
}
\makeatother

\begin{document}
\title{\rev{Dimensional reduction for } optical beams with thermal nonlocal \rev{nonlinearity}}
\author{F. Lorenzi$^{1, *}$ and L. Salasnich$^{2,3}$}
\affiliation{
$^{1}$Dipartimento di Ingegneria dell'Informazione, Universit\`a di Padova, Via Gradenigo 6A, 35131 Padova, Italy\\
$^{2}$Dipartimento di Fisica e Astronomia "Galileo Galilei", Universit\`a di Padova, Via Marzolo 8, 35131 Padova, Italy\\
$^{3}$Istituto Nazionale di Fisica Nucleare (INFN), Sezione di Padova, Via Marzolo 8, 35131 Padova, Italy\\
}

\begin{abstract}
Nonlocal optical nonlinearities arising from the thermorefractive effect provide a long-range material response determined by heat diffusion and absorption. In graded-index media, this nonlocality fundamentally alters modal interactions, yet its accurate modeling remains computationally demanding when starting from the full spatial nonlinear Schr\"odinger equation.
In this work, \rev{inspired by the nonpolynomial Schr\"odinger equation (NPSE) framework, } we extend the \rev{dimensional reduction techniques } to incorporate thermally mediated nonlocal nonlinearities. By coupling the optical field to an equation for the temperature-induced refractive index change, and employing a variational ansatz based on Laguerre–Gauss modes of the annular kind, of arbitrary azimuthal order, we derive explicit analytic expressions for the variational equations.
The resulting effective model captures the dependence of the nonlinear interaction on mode order and degree of nonlocality, providing a tractable reduced description of \rev{the } dynamics in thermal nonlocal media.
\end{abstract}

\maketitle

\section{Introduction}

Nonlocal optical nonlinearities represent an interesting self-interaction modality for waves propagating in media. They are based on the fact that the material optical response at a specific point is determined by the combined effect of the wave over an extended region of space \cite{contiObservationOpticalSpatial2004}. This nonlocality drastically alters the dynamics of the optical field compared to local nonlinearity like the Kerr effect, often acting as a stabilizing mechanism for nonlinear waves. It has indeed been extensively predicted and experimentally demonstrated that nonlocal nonlinearities can arrest catastrophic wave collapse \cite{bangCollapseArrestSoliton2002a, krolikowskiModulationalInstabilitySolitons2004, turitsynSpatialDispersionNonlinearity1985}, suppress modulational instability in defocusing regimes \cite{krolikowskiModulationalInstabilityNonlocal2001}, and support the formation of stable spatial solitary waves \cite{suterStabilizationTransverseSolitary1993, contiOpticalSpatialSolitons2005}. Beyond stabilization, the nonlocal response profoundly affects soliton mobility, particularly in the presence of optical lattices \cite{xuSolitonMobilityNonlocal2005}. Nonlocal interactions also arise in other systems, notably dipolar Bose–Einstein condensates \cite{gligoricCollapseInstabilitySolitons2009, gammalCriticalNumbersAttractive2002a}, where the long-range dipolar forces enable the emergence of supersolid states of matter \cite{tanziObservationDipolarQuantum2019}.

Among the mechanisms generating optical nonlocality, thermal nonlinearities are especially important due to their ubiquity in absorptive media and their long-range character set by heat diffusion. Originating from light absorption and subsequent temperature redistribution \cite{litvakSelffocusingPowerfulLight1966, brochardThermalNonlinearRefraction1997, sheldonLaserinducedThermalLens1982}, the thermal response is governed by a Poisson-like equation whose source is the absorbed optical power. Unlike rapidly decaying nonlocal responses, thermal nonlinearity is strongly influenced by geometry: boundary conditions shape the refractive-index profile and allow controlling soliton trajectories \cite{rotschildSolitonsNonlinearMedia2005, minovichExperimentalReconstructionNonlocal2007}. This leads to a variety of nonlinear phenomena parametrized by the geometry of the medium \cite{alfassiBoundaryForceEffects2007a, alberucciPropagationOpticalSpatial2007a, alberucciSpatialOpticalSolitons2015, alberucciNonlinearBouncingNonlocal2007a}. Thermal diffusion can also sustain dispersive spatial shock waves \cite{ghofranihaShocksNonlocalMedia2007a, marcucciOpticalSpatialShock2019}, offering a platform for studying extreme wave dynamics.
A crucial feature of thermal nonlocality is its ability to stabilize nonlinear beams and spatial solitons. In local focusing cubic media, beams carrying orbital angular momentum undergo azimuthal instabilities and break into filaments \cite{bigelowBreakupRingBeams2004}. Thermal diffusion suppresses this breakup, enabling stable vortex beams and ring solitons \cite{yakimenkoStableVortexSolitons2005a, briedisRingVortexSolitons2005}. This stabilization also influences vortex nucleation and the dynamics of vortex lines \cite{biloshytskyiVortexNucleationNonlocal2019}.
Parallel to these developments, interest in multimode optical fiber dynamics, and in particular in graded-index fibers (GRIN), has grown, motivated by the demand for higher data capacity in optical communications and the exploration of new nonlinear regimes \cite{polettiDescriptionUltrashortPulse2008, antikainenFateSolitonHigh2019}. GRIN fibers with parabolic index profile have \rev{the additional } advantage \rev{of being } analytically treatable \cite{karlssonDynamicsSelffocusingSelfphase1992, wabnitz, shtyrinaCoexistenceCollapseStable2018, aguilar-cardosoTailoringSpatialModes2025b}. Recent work has shown that interacting Laguerre–Gauss beams in GRIN fibers exhibit modified self-focusing and distinct critical powers compared with single-mode excitation \cite{saSelffocusingMultipleInteracting2019}.

However, studying nonlocal thermal nonlinearities in a generic GRIN medium poses significant computational challenges, since fully spatial numerical simulations are costly, therefore, reduced dimensional models are desirable. Recently, we introduced a nonpolynomial Schr\"odinger equation (NPSE) \cite{salasnich2002a, salasnichEffectiveWaveEquations2002a} for multimode fibers via a variational reduction of the full spatial nonlinear Schr\"odinger equation \cite{lorenzi-variational-2025}. While effective for local Kerr nonlinearities, that model did not include nonlocal interactions. In this article, we discuss the extension of the NPSE framework to nonlocal thermal nonlinearities, deriving a coupled model consisting of a full spatial NLSE and a Poisson-type equation for the thermally induced index change. Using a variational ansatz based on Laguerre–Gauss modes with arbitrary azimuthal index and zero radial index \cite{karimiLostFoundRadial2014}, we obtain explicit expressions for the nonlocal contribution. We show that the standard NPSE ansatz is however insufficient to describe diffraction properly, and a radial phase term becomes necessary. The resulting effective one-dimensional model captures the dependence of the nonlinear interaction on mode order and transverse width, providing a generalized framework for thermally mediated interactions.

\section{Fundamental equations}
\subsection{Spatial nonlinear Schr\"odinger equation in a graded-index medium}
We consider the monochromatic electric field envelope $\Phi(x,y,z)$ propagating in an isotropic medium with refractive index $n_0(x,y)$. 
Under the paraxial approximation, the propagation equation reads \cite{lorenzi-variational-2025, wabnitz}
\begin{align}
    i\partial_z \Phi &= -i\frac{\alpha_0}{2}\Phi -\frac{1}{2}\nabla_\perp^2 \Phi 
\nonumber \\ 
&+ W(x,y)\Phi + g\, n_{\mathrm{\rm non-loc}}(x,y,z)\,\Phi ,
\label{eq:nlse-3d}
\end{align}
where linear losses are modeled through the absorption coefficient $\alpha_0$; $W(x,y)$ represents the GRIN potential,
\begin{equation}
W(x,y) = \frac{1}{2}\Omega^2 r^2 ,
\end{equation}
with $r^2 = x^2+y^2$, and the nonlinear index shift is written as
\begin{equation}\label{eq:shift}
n_{\mathrm{\rm non-loc}}(x,y,z) = \frac{\Delta n(x,y,z)}{n_0}.
\end{equation}
with $n_0$ a background refractive index.
Unlike in \cite{lorenzi-variational-2025}, we consider a monochromatic field, neglecting the time dependence of the field $\Phi$. The normalization units in Eq.~\ref{eq:nlse-3d} can be conveniently chosen as the beam waist $w_0$ on the transverse dimensions $(x, y)$ and the corresponding Rayleigh range $z_R=\pi w_0^2/\lambda_0$ \cite{saleh} on the axial dimension $z$, $\lambda_0$ being the wavelength associated to a bulk material with the refractive index present at the propagation axis. 


\subsection{Steady-state temperature field}
The nonlinear shift in the refractive index \eqref{eq:shift} is given in our model by the thermo-optical effect, namely the refractive index change induced by a change of temperature of the optical medium. The change of temperature is itself induced by the local absorption, but due to thermal diffusion it will be nonlocal in space at the steady state.
The field $\mathcal{T}(x,y,z)$ describing temperature variation with respect to a reference ambient temperature, obeys the following steady-state heat equation \cite{marcucciOpticalSpatialShock2019, ghofranihaShocksNonlocalMedia2007a}
\begin{equation}\label{eq:poisson}
-\nabla^2 \mathcal{T} = \eta\, |\Phi|^2 ,
\end{equation}
where $\eta$ depends on the absorption $\alpha_0$ and the thermal diffusivity $\kappa$ through
\begin{equation}
\eta = \frac{\alpha_0}{\rho c_p \kappa},
\end{equation}
with $\rho$ the material density and $c_p$ the specific heat at constant pressure.
We will assume that the thermodynamic parameters $\rho$, $c_p$ and $\kappa$ are temperature independent, since we assume to have overall weak temperature variations, so we can use the parameters evaluated at a fixed ambient temperature and pressure.
The stationary thermal field is solved in \rev{the } presence of the following boundary conditions: over the transverse directions $x$ and $y$ we have null boundary conditions at infinity, and over the longitudinal direction $z$ we consider null boundary conditions at a flat input facet at $z=0$, and at a flat output facet at $z=L$.
The refractive index change associated with the temperature variation is given by the local expression
\begin{equation}
\Delta n(x,y,z) = \beta\, \mathcal{T}(x,y,z),
\end{equation}
where $\beta = dn/d\mathcal{T}$ is the linear thermo-optic coefficient, which can be positive or negative, depending on the material.

In \rev{the } presence of an optical beam propagating in the medium, $\mathcal{T}$ varies slowly along $z$ and the 
axial component of the Laplacian in Eq.~\eqref{eq:poisson} may be replaced by an effective confinement parameter $1/L_{\mathrm{eff}}^2$ \cite{marcucciOpticalSpatialShock2019, contiOpticalSpatialSolitons2005, contiObservationOpticalSpatial2004, alberucciPropagationSpatialOptical2010, alberucciSpatialOpticalSolitons2015}. 
A convenient way to understand this approximation is to recall that the stationary temperature profile along $z$ is expressed by the eigenmodes of the axial Laplacian, with the boundary conditions at $z=0$ and $z=L$ (see also \cite{alberucciPropagationSpatialOptical2010}).
When analyzing the temperature profile distant from the edges, in the weak absorption regime, the full three-dimensional Green function expressed in Refs.~\cite[Eq.~(13)]{alberucciPropagationSpatialOptical2010}, \cite[Eq.~(5)]{alberucciSpatialOpticalSolitons2015} can be approximated so \rev{as } to contain only the contribution from the fundamental mode along the $z$ direction.
The result of this approximation is that $L_{\mathrm{eff}}=L/\pi$. The case of strong absorption can be handled using a similar approach but with a modified value of $L_{\mathrm{eff}}$ \cite{ghofranihaShocksNonlocalMedia2007a}.
The resulting equation, valid sufficiently far from the sample boundaries, is therefore a 2D screened Poisson-type equation
\begin{equation}
\left(-\nabla_\perp^2 + \frac{1}{L_{\mathrm{eff}}^2}\right) \mathcal{T}(x,y,z)
= \eta\, |\Phi(x,y,z)|^2 .
\label{eq:screened-poisson}
\end{equation}
The formal solution of Eq.~\eqref{eq:screened-poisson} at fixed $z$ is the transverse convolution
\begin{equation}
\mathcal{T}(x, y, z) = 
\eta \int d^2\mathbf{r}_\perp' \,
G(\mathbf{r}_\perp-\mathbf{r}'_\perp)\, |\Phi(\mathbf{r}'_\perp,z)|^2,
\label{eq:thermal-convolution}
\end{equation}
where $G$ is the Green's function, and $\mathbf{r}_\perp$, $\mathbf{r}_\perp'$ indicate transverse coordinates. In two transverse dimensions, this Green function is known in closed 
form and involves the modified Bessel function $K_0$ \cite{abramowitz,yakimenkoStableVortexSolitons2005a}:
\begin{equation}
G(\mathbf{r}) = \frac{1}{2\pi} K_0\!\left(\frac{r}{L_{\mathrm{eff}}}\right).
\end{equation}
At short distances $K_0$ reproduces the logarithmic singularity 
of the 2D Poisson kernel, while for $r \gg L_{\mathrm{eff}}$ it decays exponentially, 
reflecting the suppression of long-range diffusion by the heat transfer on the facets.

Within this model, the variation of the nonlinear index will satisfy
\begin{equation}
g\,n_{\mathrm{non-loc}}(\mathbf{r},z)
= \frac{\gamma}{2\pi} 
\int d^2\mathbf{r}_\perp' \,
K_0\!\left(\frac{|\mathbf{r}_\perp-\mathbf{r}'_\perp|}{L_{\mathrm{eff}}}\right)
|\Phi(\mathbf{r}'_\perp,z)|^2,
\label{eq:nonlocal-kernel}
\end{equation}
with $\gamma = g\beta \alpha_0/(\rho c_p \kappa)$.

By combining the above elements, the optical propagation obeys the full spatial nonlocal NLSE
\begin{equation}
i\partial_z \Phi = -\rev{i}\tfrac{\alpha_0}{2}\Phi -\tfrac12\nabla_\perp^2 \Phi 
+ W(r)\Phi 
+ \gamma\, \Theta[|\Phi|^2]\,\Phi ,
\label{eq:nonlocal-nlse}
\end{equation}
where the nonlocal operator is defined through the convolution
\begin{equation}\label{eq:theta}
\Theta[|\Phi|^2](\mathbf{r}_\perp,z)
= \frac{1}{2\pi} \int d^2\mathbf{r}'_\perp\,
K_0\!\left(\frac{|\mathbf{r}_\perp-\mathbf{r}'_\perp|}{L_{\mathrm{eff}}}\right)
\bigl|\Phi(\mathbf{r}'_\perp,z)\bigr|^2 \rev{\,.}
\end{equation}
We remark that the 2D screened-Poisson model offers a substantial computational advantage over the full diffusion equation, as it integrates naturally into beam-propagation simulations where the nonlocal response is updated independently at each axial step. It also provides a convenient starting point for the variational dimensional reduction introduced below.
Such a model is not unique to thermal media: screened-Poisson responses also describe reorientation nonlinearity in liquid crystals \cite{alberucciSpatialOpticalSolitons2015} and electrostrictive effects \cite{contiOpticalSpatialSolitons2005} and electron-ion interactions in plasmas \cite{yakimenkoStableVortexSolitons2005a}. In all these cases, the stationary transverse index profile is well approximated by a screened Poisson equation \cite{ghofranihaShocksNonlocalMedia2007a}.

\section{Effective action with thermal nonlocality}

In order to \rev{perform a dimensional reduction, inspired by the NPSE formalism, in the presence of }  thermal nonlocality, it is convenient to rewrite
Eq.~\eqref{eq:nonlocal-nlse} in variational form. 
The lossless version of Eq.~\eqref{eq:nonlocal-nlse} (i.e. setting $\alpha_0=0$), is the equation of motion associated with the action
\begin{equation}
S = \int dz \int d^2\mathbf{r}_\perp\; \mathcal{L},
\end{equation}
with Lagrangian density
\begin{align}
\mathcal{L} &=
\frac{i}{2}\bigl(\Phi^* \partial_z \Phi - \Phi\, \partial_z \Phi^* \bigr)
- \frac{1}{2} |\nabla_\perp \Phi|^2
- W(r)\,|\Phi|^2
\nonumber\\
&\quad
- \frac{\gamma}{2}\, |\Phi|^2\,
\Theta[|\Phi|^2].
\label{eq:full-3d-lagrangian-no-time}
\end{align}
The first line of Eq.~\eqref{eq:full-3d-lagrangian-no-time} contains the local
contributions due to diffraction and the GRIN potential, while the second line
encodes the thermorefractive nonlocality through a nonlocal quartic term.
With this choice, the Euler--Lagrange equation is Eq.~\eqref{eq:nonlocal-nlse}, with the exception of losses, which we include phenomenologically a posteriori for consistency with the source of nonlocality.
%
We assume the transverse field is a single Laguerre--Gauss (LG) mode with radial index $p$ and azimuthal index $m$, and $S=|m|$,
\begin{equation}\label{eq:lgansatz}
  \Phi(r,\theta,z)
  = A(z)\,T_{pS}\bigl(r;\sigma(z)\bigr)\,
    \exp\!\left[i\,\frac{b(z)}{2}\,r^{2}\right]\,
    e^{i m\theta},
\end{equation}
where $\sigma(z)$ is a variational transverse width and
$T_{pS}$ is normalized as
\begin{equation}
\int_0^\infty dr\, r\, |T_{pS}(r;\sigma)|^2 = 1.
\end{equation}
\rev{T}his implies that the power of the beam is simply expressed as $P=2\pi|A|^2$.
The parameter $b(z)$ regulates the phase profile over the radial direction. It is motivated by the known phase profile of paraxial beams, and was used in past literature for variational study of beam propagation \cite{karlssonDynamicsSelffocusingSelfphase1992}.
The LG transverse profile is written as
\begin{equation}
T_{pS}(r,\sigma) =
\frac{1}{\sigma}\,
C_{pS}\,
\left(\frac{r}{\sigma}\right)^{S}
L_p^{S}\!\left(\frac{r^2}{\sigma^2}\right)
\exp\!\left[-\frac{r^2}{2\sigma^2}\right],
\end{equation}
with
\begin{equation}
C_{pS} =
\sqrt{\frac{2p!}{(p+S)!}} \rev{\,,}
\end{equation}
\rev{where } $p$ is the radial number.


\subsection{Effective Lagrangian: local terms}

Inserting the ansatz into the full Lagrangian density and 
integrating over the transverse plane yields
\begin{align}
  L_{\mathrm{loc}}
  =& i A^* \dot{A} \nonumber\\
    &- \frac{\xi_{pS}}{2} \bigl[\sigma^{-2} + (\Omega^2 + b^2+\dot{b}) \sigma^2\bigr]\,
      |A|^2 \rev{\,,} \nonumber\\ 
  \label{eq:LL-general}
\end{align}
where the coefficients are
\begin{equation}
\xi_{pS}=S+2p+1 ,
\end{equation}
as in Ref.~\cite{lorenzi-variational-2025}. The dot notation indicates differentiation with respect to the propagation variable $z$.
These terms are identical to the local NPSE case \cite{lorenzi-variational-2025, lorenzi-atomic-2024}, except for the presence of the chirp term.

\subsection{Effective action: nonlocal term}

The thermal nonlocality enters the propagation Lagrangian in 1D through the last term
of the Lagrangian density~\eqref{eq:full-3d-lagrangian-no-time},
\begin{equation}
\mathcal{L}_{\rm non-loc}
=
- \frac{\gamma}{2}\,
|\Phi(\mathbf{r}_\perp,z)|^2\,
\Theta[|\Phi|^2](\mathbf{r}_\perp,z),
\end{equation}
where the nonlocal thermal potential is defined as the convolution \eqref{eq:theta}.
The corresponding contribution to the effective one-dimensional Lagrangian
$L(z)$ is obtained by integrating over the transverse plane:
\begin{align}
L_{\rm non-loc}
&= \int d^2\mathbf{r}_\perp\; \mathcal{L}_{\rm non-loc}
\nonumber\\
&=
- \frac{\gamma}{4\pi}
\int d^2\mathbf{r}_\perp\, |\Phi(\mathbf{r}_\perp, z)|^2 \nonumber\\
&\times\int d^2\mathbf{r}_\perp'\,
K_0\!\left(\frac{|\mathbf{r}_\perp-\mathbf{r}_\perp'|}{L_{\rm eff}}\right)
|\Phi(\mathbf{r}_\perp', z)|^2  \,,
\end{align}
we remark that this expression is symmetric under the exchange of $\mathbf{r}_\perp$ and $\mathbf{r}_\perp'$.
This is the general structure of the thermal nonlocal interaction at the Lagrangian level; using the LG ansatz \eqref{eq:lgansatz} the nonlocal contribution becomes purely quartic in the axial amplitude $A$:
\begin{equation}
L_{\rm non-loc}
=
- \frac{\gamma}{2}\,|A|^4\,
\mathcal{I}_{pS}(\sigma),
\label{eq:LNL-general}
\end{equation}
where the transverse nonlocal integral is
\begin{align}
\mathcal{I}_{pS}(\sigma)
=& \frac{1}{2\pi}\int d^2\mathbf{r}_\perp\, d^2\mathbf{r}_\perp'\,
K_0\!\left(\frac{|\mathbf{r}_\perp-\mathbf{r}_\perp'|}{L_{\rm eff}}\right) \nonumber \\
&\times |T_{pS}(r;\sigma)|^2
|T_{pS}(r';\sigma)|^2 .
\label{eq:Ips-definition}
\end{align}
The dependence on the mode indices $(p,S)$ and on the width $\sigma$
enters only through $\mathcal{I}_{pS}(\sigma)$, which is the quantity
to be evaluated explicitly.

We consider the nonlocal integral
\begin{align}
\mathcal{I}_{pS}(\sigma)
=& \frac{1}{2\pi}
\int d^2\mathbf{r}_\perp\, d^2\mathbf{r}_\perp'\,
K_0\!\left(
\frac{|\mathbf{r}_\perp - \mathbf{r}_\perp'|}{L_{\rm eff}}
\right) \nonumber \\
&\times
\rho_{pS}(r;\sigma)\,
\rho_{pS}(r';\sigma),
\label{eq:Ipm-double}
\end{align}
where $\rho_{pS}(r;\sigma)=|T_{pS}(r;\sigma)|^2$ is radially symmetric.
The integral can be simplified by using the Fourier convolution theorem in two dimensions (see Appendix A)
\begin{equation}
\mathcal{I}_{pS}(\sigma)
=
2\pi
\int_0^\infty dq\, q\,
\frac{L_{\rm eff}^2}{1+L_{\rm eff}^2 q^2}\,
\tilde\rho_{pS}(q\rev{;\sigma})^2
.
\label{eq:Ipm-final-correct}
\end{equation}
\rev{where 
\begin{equation}
\tilde\rho_{pS}(q; \sigma)
=
\int_0^\infty dr\, r\,
\rho_{pS}(r;\sigma)\,J_0(qr),
\label{eq:rho-Hankel}
\end{equation}
is the Hankel transform of $\rho$.}

The effective Lagrangian in this case is obtained combining the local part Eq.~\eqref{eq:LL-general} with the nonlocal part Eq.~\eqref{eq:LNL-general}, with the simplified value of the nonlocal contribution expressed implicitly by Eq.~\eqref{eq:Ipm-final-correct}.
The Euler--Lagrange equation for the axial amplitude follows from
$\partial L/\partial A^*=0$, since $L$ does not depend on
$\dot{A}^*$. 
To have a consistent account of the thermal effects, we add the absorption in the equation of motion, \rev{taking} into account the transverse shape of the mode
\begin{align}
  i\,\dot{A}
  =& -\,i\frac{\alpha_0}{2}A\nonumber \\
    &+ \frac{\xi_{pS}}{2}
      \bigl[\sigma^{-2} + (\Omega^2 + b^2+\dot{b}) 
\sigma^2\bigl]\,A \nonumber \\
    &+ \gamma \mathcal{I}_{pS}(\sigma)\,|A|^2 A.
  \label{eq:A_eom_general}
\end{align}
The first two terms correspond to the local contributions,
while the third term is the nonlocal thermo--refractive correction.
The transverse width $\sigma(z)$ enters the Lagrangian algebraically,
so its equation of motion reduces to the condition
$\partial L / \partial \sigma = 0$.
The local part yields
\begin{equation}
\frac{\partial L_{\mathrm{loc}}}{\partial\sigma}
=
\xi_{pS} |A|^{2}\left(
\sigma^{-3}-(\Omega^2+b^2+\dot{b})\sigma
\right),
\label{eq:dLlocds}
\end{equation}
while the nonlocal part gives
\begin{equation}
\frac{\partial L_{\mathrm{nonloc}}}{\partial\sigma}
=
-\frac{\gamma}{2}|A|^{4}
\frac{d \mathcal{I}_{pS}}{d\sigma}.
\label{eq:dLnonlocds}
\end{equation}
Collecting all terms, the condition $\partial L/\partial\sigma = 0$
yields the algebraic equation
%
\begin{align}
  &\xi_{pS}|A|^2
  \left[
    \sigma^{-3}
    - \left(\Omega^2 + b^2 + \dot{b}\right)\sigma
  \right] \nonumber \\
  &-\frac{\gamma}{2}|A|^{4}
\frac{d \mathcal{I}_{pS}}{d\sigma}
  = 0.
  \label{eq:sigma_eom_general}
\end{align}
Finally the Euler-Lagrange equation for the radial phase chirp parameter is
\begin{equation}
  \frac{d}{dz}\bigl(\sigma^2 |A|^2\bigr)
  = 2\,b\,\sigma^2\,|A|^2.
  \label{eq:b_eom}
\end{equation}
Equations \eqref{eq:A_eom_general}, \eqref{eq:sigma_eom_general} and \eqref{eq:b_eom}
define a dimensional reduction: the axial field evolves under a $\sigma$--dependent nonlinearity, while the transverse width is determined by the balance between the local and nonlocal thermal contributions.
\rev{These equations constitute the main result of the work. While they are not immediately written with an algebraic nonlinearity, \rev{similarly to } the NPSE approach \cite{salasnich2002a}, closed forms are available for specific choices of the transverse distributions, like the Gaussian one. }
We remark that, unlike the previously studied systems, where it was possible to express explicitly the value of $\sigma$ and substitute it back into the equation for the axial amplitude, in this case, since $\mathcal{I}_{pS}$ is a complicated function of $\sigma$, this is no \rev{longer } possible. However, it is remarkable that closed-form expressions can be found for $\mathcal{I}_{0S}$, i.e. the case with $p=0$, as shown in Appendix B. 

\subsection{Special case: Gaussian beam}
For the fundamental Gaussian mode ($p=0$, $m=0$, hence $S=0$), the
transverse profile reduces to
\begin{equation}
T_{00}(r;\sigma) = \frac{\sqrt{2}}{\sigma}
\exp\!\left(-\frac{r^2}{2\sigma^2}\right),
\end{equation}
so that the radial intensity is
\begin{equation}
\rho_{00}(r;\sigma) = |T_{00}(r;\sigma)|^2
= \frac{2}{\sigma^2}\,
\exp\!\left(-\frac{r^2}{\sigma^2}\right).
\end{equation}
Its Hankel transform is again a Gaussian,
\begin{equation}
\widetilde{\rho}_{00}(q)
= e^{-\sigma^2 q^2/4}.
\end{equation}
Inserting this into Eq.~\eqref{eq:Ipm-final-correct} yields
\begin{equation}
\mathcal{I}_{00}(\sigma)
=
2\pi L_{\rm eff}^2
\int_0^\infty dq\,
\frac{q\, e^{-\sigma^2 q^2/2}}
{1+L_{\rm eff}^2 q^2}.
\end{equation}
This integral can be performed in terms of the exponential integral $\operatorname{Ei}$ \cite{abramowitz} %
\begin{equation}\label{eq:I00-result}
\mathcal{I}_{00}(\sigma)
=
-\pi \exp\!\left(
\frac{\sigma^{2}}{2L_{\mathrm{eff}}^{2}}
\right)
\operatorname{Ei}\left(
-\frac{\sigma^{2}}{2L_{\mathrm{eff}}^{2}}
\right),    
\end{equation}
its derivative reads
\begin{equation}
\frac{d \mathcal{I}_{00}}{d\sigma}
=
\frac{\sigma}{L_{\mathrm{eff}}^{2}}\,\mathcal{I}_{00}(\sigma)
-
\frac{2\pi}{\sigma}.
\label{eq:dIds_gaussian}
\end{equation}
The corresponding integrals for $S\geq1$ are less straightforward and can be obtained as discussed in Appendix B.

We remark that, if the thermal nonlocal term \eqref{eq:I00-result} is considered in the limit of $L_{\mathrm{eff}}\to 0$, the equations for the beam evolution reduce to the ones derived in the variational model of Ref.~\cite{karlssonDynamicsSelffocusingSelfphase1992} (the limit is also discussed in the derivation of \cite[Eq.~(7)]{ghofranihaShocksNonlocalMedia2007a}). Unlike the treatment in Ref.~\cite{karlssonDynamicsSelffocusingSelfphase1992}, in the present model it is not possible to simply generalize the results to pulses, as the response time of the thermal properties would make the analysis much more involved, even neglecting chromatic dispersion.
Finally, in the special case of a linear evolution, Eq.~\eqref{eq:sigma_eom_general} becomes the well-known equation for beam width evolution (see also Ref.~\cite{lewisClassicalQuantumSystems1967}).

\begin{figure}[t]
\centering
\includegraphics[width=\columnwidth]{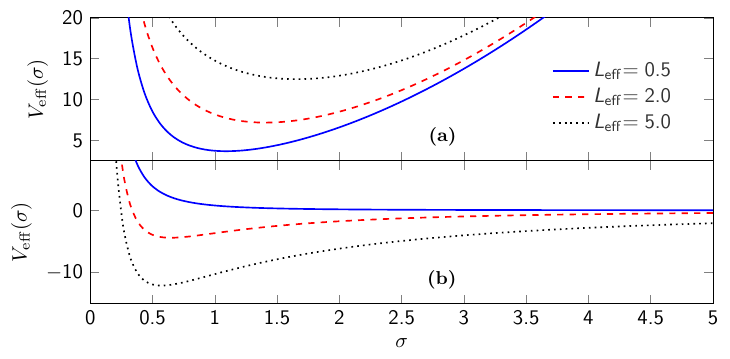}
\caption{Effective potential $V_{\mathrm{eff}}(\sigma)$ of the variational
width dynamics for different nonlocality lengths $L_{\mathrm{eff}}$.
Panel~(a) shows the defocusing regime $\gamma=0.5$ in the absence of a confining potential ($\Omega=0$), and panel~(b) shows the focusing
regime $\gamma=-0.5$ in the presence of a confining potential $\Omega=1$. The index $S$ is chosen to $S=2$.}
\label{fig:potential}
\end{figure}

\section{Numerical simulations}
\begin{figure}[t]
\centering
\includegraphics[width=\columnwidth]{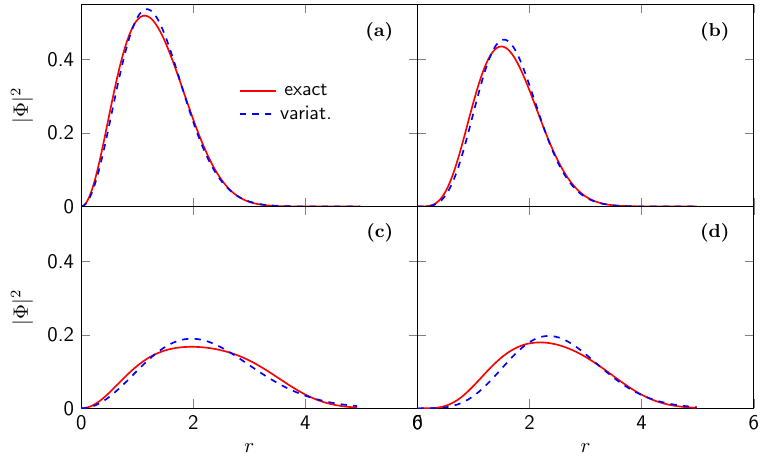}
\caption{Comparison of stationary radial intensity profiles of the beam $|\Phi(\mathbf{r})|^2$ obtained from imaginary time evolution of the paraxial diffraction operator (solid red lines) and the variational ansatz (dashed blue lines). Panels (a) and (b) show the case of $L_{\mathrm{eff}} = 0.5$ for vortex indices $S=1$ and $S=2$, respectively. Panels (c) and (d) show the regime of $L_{\mathrm{eff}} = 5.0$ for $S=1$ and $S=2$. In this plot $\Omega=1$, and $\gamma=0.5$.}
\label{fig:stationary_grid_repulsive}
\end{figure}

\begin{figure}[t]
\centering
\includegraphics[width=\columnwidth]{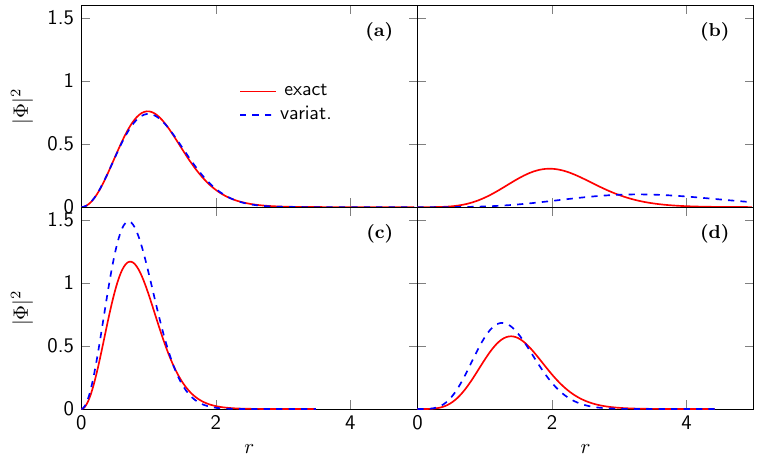}
\caption{Comparison of stationary radial intensity profiles of the beam
$|\Phi(\mathbf{r})|^{2}$ obtained from imaginary time evolution of the
paraxial diffraction operator (solid red lines) and the variational
ansatz (dashed blue lines). Panels (a) and (b) show the case of ($L_{\mathrm{eff}} = 1.2$) for vortex indices $S=1$ and $S=2$, respectively. Panels (c) and (d) show the case of $L_{\mathrm{eff}} = 5.0$, also for $S=1$ and $S=2$. In this plot $\Omega=0$ and $\gamma=-0.22$.}
\label{fig:stationary_grid_attractive}
\end{figure}

To validate the reduced model, we performed direct simulations of the full two-dimensional paraxial diffraction equation with thermal nonlocality using a standard split-step Fourier method. The linear propagation was computed in the transverse momentum domain, while the nonlinear thermorefractive potential was updated at each axial step through convolution with the kernel discussed in Eq.~\eqref{eq:nonlocal-kernel}. This approach follows established numerical strategies for nonlocal nonlinear Schrödinger equations~\cite{krolikowskiModulationalInstabilityNonlocal2001, yakimenkoStableVortexSolitons2005a}, and allows for accurate benchmarking of beam evolution, including dynamical regimes where the dimensionally reduced model is expected to perform poorly, like the ones of dispersive shock waves in the self-defocusing regime \cite{ghofranihaShocksNonlocalMedia2007a}, and symmetry-breaking azimuthal instability \cite{yakimenkoStableVortexSolitons2005a}. The reduced variational model represented by Eqs.~(\ref{eq:A_eom_general}, \ref{eq:sigma_eom_general}, \ref{eq:b_eom}) is solved using a standard fourth-order Runge-Kutta method.

\subsection{Stationary beam profiles}
Starting from the equations of motion, we consider the conservative case
$\alpha_0=0$. In this regime Eq.~\eqref{eq:A_eom_general} imposes that the beam power $P = 2\pi |A|^{2}$ is conserved. Using this information and substituting in Eq.~\eqref{eq:b_eom}, we obtain that the derivative $\dot{\sigma}$ is proportional to $b$: to have a stationary value, we require $b=0$. A particularly interesting analysis is then to solve Eq.~\eqref{eq:sigma_eom_general} for the stationary value of $\sigma$. In the following, we set $|A|^2=1$. The aforementioned Eq.~\eqref{eq:sigma_eom_general} in the static case can be interpreted as the condition of stationarity of the following effective potential
\begin{equation}
  V_{\mathrm{eff}}(\sigma)
  = \frac{\xi_{pS}}{2}\!\left(\frac{1}{\sigma^{2}}
  + \Omega^{2}\sigma^{2}\right)
  + \frac{\gamma |A|^2}{2}\,\mathcal{I}_{0S}\!\bigl(\sigma\bigr)\rev{\,.}
  \label{eq:Veff_sigma}
\end{equation}

The structure of the effective potential $V_{\mathrm{eff}}(\sigma)$ changes
significantly with the degree of thermal nonlocality, as shown in
Fig.~\ref{fig:potential}. In the defocusing regime
(Fig.~\ref{fig:potential}(a)), increasing $L_{\mathrm{eff}}$ shifts the
minimum of the potential toward larger widths,
indicating weaker transverse confinement and a broader stationary beam. In the
focusing regime (Fig.~\ref{fig:potential}(b)), the nonlocal contribution
changes the potential more dramatically: the attractive term induced by
diffusion becomes important at small $\sigma$, and the depth of the potential
well increases with $L_{\mathrm{eff}}$.

To assess the validity of the dimensional reduction, we first compare the stationary states obtained via imaginary time propagation 
\cite{jacksonVorticesTrappedBose2000, tikhonenkovVortexSolitonsDipolar2008,
baoComputingGroundState2004a}
with the predictions of the variational method.
Figure~\ref{fig:stationary_grid_repulsive} compares the radial intensity
profiles $|\Phi(\mathbf{r})|^2$ in a defocusing regime, $\gamma = 0.5$, with a
graded-index potential, $\Omega = 1$. In the case of short-range nonlocality of panels (a) and (b), having $L_{\mathrm{eff}} = 0.5$, the variational prediction shows excellent agreement with the
exact numerical solutions for both vortex orders $S = 1$ and
$S = 2$. However, as the thermal response becomes more nonlocal (panels (c) and (d), with $L_{\mathrm{eff}} = 5.0$), the physical beam profile deforms and departs from
the functional form assumed in the variational ansatz.
A similar analysis is done in the focusing case and without confining potential in
Fig.~\ref{fig:stationary_grid_attractive}, with a parameter
$\gamma = -0.22$. In this analysis the nonlinear parameter and the effective length $L_{\mathrm{eff}}$ were chosen to prevent the solver from collapsing to a filamented state. \rev{Among } the simulations with $L_{\mathrm{eff}}=1.2$ (panels (a) and (b)), the one with high vorticity in panel (b) displays a poor matching of the variational results. This is interpreted by considering the fact that this choice of parameters leads to a situation similar to the one represented in panel (b) of Figure~\ref{fig:potential}: when diminishing the effective length, the curvature of the effective potential diminishes more and more. This makes the \rev{system }  much more \rev{sensitive } to small variations in the radial shape and deviations from the Laguerre--Gauss shape. The variational model still reproduces the qualitative broadening of the beam in exact solutions in the strongly nonlocal regime $L_{\mathrm{eff}} = 5.0$ (panels (c) and (d)).

\subsection{Beam propagation}
\begin{figure}[t]
\centering
\includegraphics[width=\columnwidth]{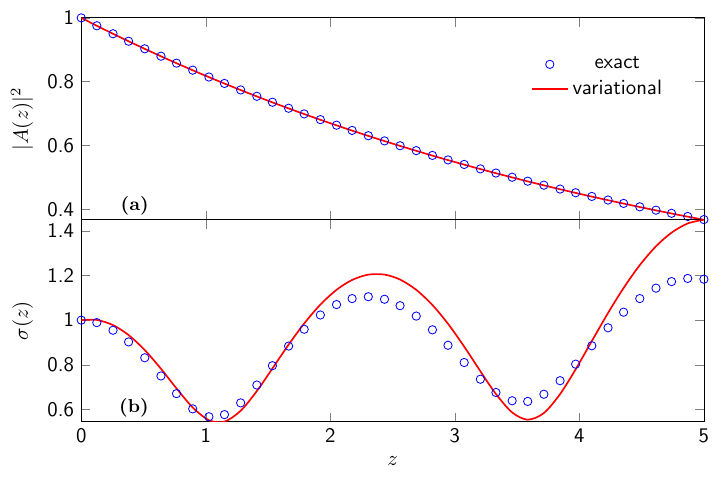}
\caption{
Comparison of the axial beam dynamics obtained from full 2D paraxial
propagation (blue markers) and the variational model (solid red
lines). Panel (a) shows the evolution of the normalized power
$|A(z)|^{2}$, while panel (b) reports the corresponding transverse width
$\sigma(z)$.
The simulations use a vortex beam with $p=0$, $S=1$ and initial
conditions $A(0)=1$, $\sigma(0)=1$, and $b(0)=0$ (constant radial phase).
The nonlinear parameters are $\gamma=-0.15$, $L_{\mathrm{eff}}=40$,
$\alpha=0.2$, and $\Omega=1$.}
\label{fig:axial_dynamics}
\end{figure}

Figure~\ref{fig:axial_dynamics} shows the dynamical evolution of a vortex beam $S = 1$ propagating with losses set to  $\alpha = 0.2$ with strong
thermal nonlocality $L_{\mathrm{eff}} = 40$. The axial amplitude $|A(z)|^2$ decays monotonically due to linear absorption (panel~(a)), while the transverse
width $\sigma(z)$ undergoes damped breathing oscillations constrained by the graded-index potential (panel~(b)). The correspondence between the reduced one-dimensional variational model (solid red lines) and the full two-dimensional split-step Fourier simulations (blue markers) is substantial. This demonstrates that, despite the moderate discrepancies observed in the static mode profiles,
the variational method accurately captures the overall beam evolution dictated by the interplay of diffraction, thermal diffusion, and the parabolic trapping potential, especially for small propagation distances.
We remark that the inclusion in the ansatz of the radial chirp term as described in Eq.~\eqref{eq:lgansatz} is important. This can be deduced  \rev{by } analyzing Eq.~\eqref{eq:sigma_eom_general}, where, in \rev{the } absence of the variable $b$, any non-dissipating beam would have a fixed width for  \rev{the entire } propagation distance.


\section{Conclusions}
In this work we have examined a reduced one-dimensional description of beam propagation in graded-index media with thermal nonlocal nonlinearities, derived from the full spatial nonlinear Schr\"odinger equation coupled to a
screened-Poisson model. By extending the variational methodology to include a nonlocal convolution term, we obtained an effective action in which the thermal interaction appears through a mode- and width-dependent nonlocal energy functional. The use of an annular Laguerre--Gauss trial function enabled closed form expressions for the Hankel transform of the intensity and yielded a compact formulation of the reduced dynamics.
The resulting evolution equations show that thermal nonlocality alters the dependence of the nonlinear interaction on the transverse width and mode order. The analysis also indicates that a single-width ansatz is insufficient to capture the dynamics in nonlocal media, and that a radial phase-chirp parameter is needed.
The variational reduction is performed for the conservative system, and linear absorption is then reintroduced phenomenologically.

The approach may be extended to multimode or partially coherent fields, where thermal diffusion mediates effective intermodal coupling. Further generalizations may \rev{include } saturable responses or time-dependent thermal relaxation, to broaden the applicability of the method to pulsed systems \rev{as well}, where the interplay with local electronic Kerr effect and chromatic dispersion would be interesting.


\section{Appendix A: Fourier-Bessel convolution}
To reduce this expression we use the Hankel transform of the kernel,
\begin{equation}
\tilde K(q)
=
\int_0^\infty dr\, r\,
K_0\!\left(\frac{r}{L_{\rm eff}}\right)\,J_0(qr),
\label{eq:Ktilde-def}
\end{equation}
which provides the inversion formula
\begin{equation}
K_0\!\left(\tfrac{|\mathbf r-\mathbf r'|}{L_{\rm eff}}\right)
=
\int_0^\infty dq\, q\, \tilde K(q)\,
J_0\!\big(q|\mathbf r-\mathbf r'|\big).
\end{equation}

Substituting this into \eqref{eq:Ipm-double} and exchanging the order of
integration yields (we drop the subscripts of $\rho$ for brevity)
\begin{equation}
\mathcal{I}_{pS}(\sigma)
=\frac{1}{2\pi}
\int_0^\infty dq\, q\,\tilde K(q)\,
J(q),
\end{equation}
\begin{equation}
J(q)
= \int d^2\mathbf r\, d^2\mathbf r'\,
\rho(r)\rho(r')\,
J_0\!\big(q|\mathbf r-\mathbf r'|\big).
\end{equation}

Using the addition theorem for Bessel functions together with
azimuthal integration \cite[\S 8.530]{gradshteyn},
\begin{equation}
\int_0^{2\pi}\!\!\!\!\! d\theta
\int_0^{2\pi}\!\!\!\!\! d\theta'\;
J_0\!\big(q|\mathbf r-\mathbf r'|\big)
=
(2\pi)^2\,
J_0(qr)\,J_0(qr'),
\end{equation}
we obtain
\begin{align}
J(q)
=&
(2\pi)^2 
\left[\int_0^\infty dr\, r\,\rho(r)\,J_0(qr)\right] \nonumber \\
&\times \left[\int_0^\infty dr'\, r'\,\rho(r')\,J_0(qr')\right].
\end{align}
\rev{Using the definition of } the Hankel transform of the radial intensity, \rev{in Eq.~\eqref{eq:rho-Hankel}, \, }
the nonlocal integral becomes
\begin{equation}
\mathcal{I}_{pS}(\sigma)
=
2\pi
\int_0^\infty dq\, q\,
\tilde K(q)\,
\bigl|\tilde\rho_{pS}(q)\bigr|^2
.
\label{eq:Ipm-Hankel-correct}
\end{equation}

Finally, inserting the explicit kernel transform
\begin{equation}
\tilde K(q)=\frac{L_{\rm eff}^2}{1+L_{\rm eff}^2 q^2},
\end{equation}
we retrieve the compact expression \eqref{eq:Ipm-final-correct}.

\section{Appendix B: closed-form Hankel transform for Laguerre-Gauss intensities}
We compute the Hankel transform $\widetilde{\rho}_{pS}(q)$ for the LG
intensity. Using the explicit expressions defining \eqref{eq:lgansatz}, one has
\begin{equation}
\rho_{pS}(r) = 
\frac{C_{pS}^{\,2}}{\sigma^{2S+2}}\,
r^{2S}\,
\left[
L_{p}^{S}\!\!\left(\frac{r^{2}}{\sigma^{2}}\right)
\right]^{2}\,
\exp\!\left(-\frac{r^{2}}{\sigma^{2}}\right),
\end{equation}
Accordingly,
\begin{equation}
\widetilde{\rho}_{pS}(q)
=
\int_{0}^{\infty}\!dr\, r\,
\rho_{pS}(r;\sigma)\, J_{0}(qr).
\end{equation}

A particularly compact expression is obtained for the vortical modes with $p=0$ and $S\neq 0$, for which
\begin{equation}
\rho_{0S}(r)
=
B_{0S}\, r^{2S}\,
\exp\!\left(-\frac{r^{2}}{\sigma^{2}}\right),
\end{equation}
with
\begin{equation}
B_{0S}
=
\frac{2}{\sigma^{2(S+1)}\,S!}.
\end{equation}
In this case the Hankel transform can be expressed using confluent hypergeometric functions
\begin{equation}
\label{eq:rhohat_vortical}
\widetilde{\rho}_{0S}(q)
= 
{}_{1}F_{1}\!\left(1+S\,;\,1\,;\,-\frac{q^{2}\sigma^{2}}{4}\right),
\end{equation}
where ${}_{1}F_{1}$ denotes the confluent hypergeometric function of the first
kind.  Equation~\eqref{eq:rhohat_vortical} provides a compact analytic
expression for the Hankel transform of the annular ($p=0$) LG vortices, and is
particularly useful when evaluating the nonlocal energy $\mathcal{I}_{0S}$
through Eq.~\eqref{eq:Ipm-final-correct}. Indeed, performing the integration for a fixed value of $S$ one finds that the quantity $\mathcal{I}_{0S}$ admits an explicit analytical expression in terms of polynomials in the variable $\sigma/L_{\mathrm{eff}}$ and expressions containing the exponential integral function $\mathrm{Ei}$. In the case $S=0$ this procedure reproduces
the closed form reported in Eq.~\eqref{eq:I00-result}. For $S>0$, the
derivative $d\mathcal{I}_{0S}/d\sigma$ required in Eq.~\eqref{eq:sigma_eom_general} is likewise available analytically once the value of $\mathcal{I}_{0S}$ is determined. 

\begin{figure}
    \centering
    \includegraphics[width=\columnwidth]{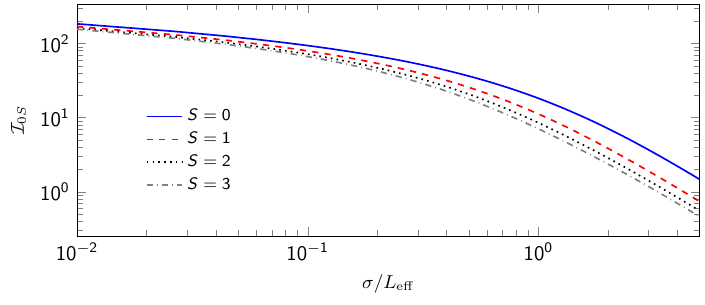}
    \caption{
    Scaling of the nonlocal interaction coefficient $\mathcal{I}_{0S}$ as a function of the normalized beam width $\sigma/L_{\mathrm{eff}}$.}
\label{fig:sigma_I0S_scaling}
\end{figure}

The scaling of the nonlocal coefficient is analyzed in
Fig.~\ref{fig:sigma_I0S_scaling}.  The $S=0$ case (solid blue line) defines an
upper bound, as increasing the azimuthal index $S$ systematically decreases $\mathcal{I}_{0S}$.
The dependence on $\sigma$ contrasts with the variational analysis obtained with local nonlinearity, where interaction energy scales as $1/\sigma^{2}$ independently of $S$. In the nonlocal case, there is no simple algebraic decay valid for all values of $\sigma$.


\bibliography{main}

\clearpage
\newpage

\end{document}